\documentclass[final,5p,times,twocolumn]{elsarticle}

\usepackage{amssymb}
\usepackage{amsmath}
\usepackage{lineno}
\usepackage{subcaption}
\journal{Nuclear Instruments and Methods A}

\begin{document}

\begin{frontmatter}


\ead{alisa.healy@cockcroft.ac.uk}
 \fntext[label2]{Work supported by the STFC core grant ST/K520133/1
}

\title{Electron-Terahertz Interaction In Dielectric-Lined Waveguide Structures For Electron Manipulation}


\author{A.L. Healy$^{a,b}$, G. Burt$^{a,b}$, S.P. Jamison$^{b,c}$}

\address{$^a$Department of Engineering, Lancaster University, LA1 4YR, UK\\
$^b$ The Cockcroft Institute, Sci-Tech Daresbury, WA4 4AD, UK\\
$^c$ Accelerator Science and Technology Centre, STFC, WA4 4AD, UK}

\begin{abstract}
Terahertz-driven dielectric-lined waveguides (DLWs) have uses in electron manipulation; in particular deflection, acceleration, and focussing. A rectangular DLW has been optimised for deflection of 100 keV electrons using a THz pulse with a centre frequency 0.5 THz. Electron-THz interaction and the effect of electron bunch injection timing on maximising deflection is presented. DLWs and corrugated waveguides are compared to discuss relative advantages and disadvantages. 

\end{abstract}


\end{frontmatter}


\section{Introduction}
\label{}
\begin{figure}[h]
	\centering
	\includegraphics[width=0.8\linewidth]{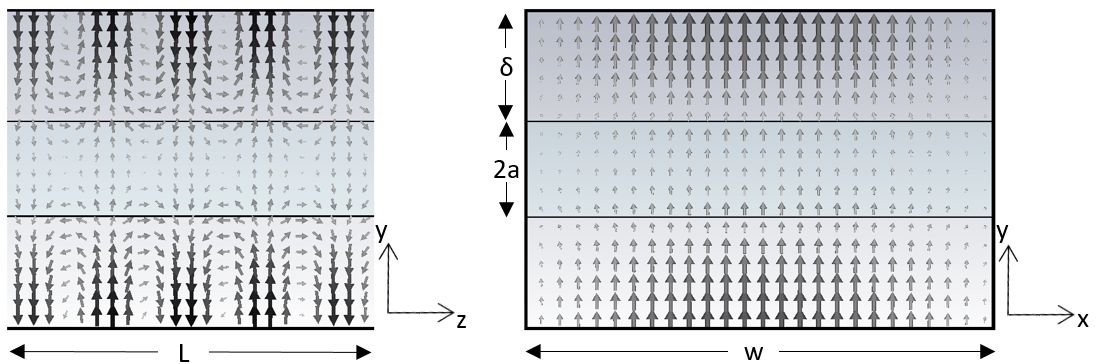}
	\caption{Cross-sectional view of the dielectric-lined waveguide, showing the LSM$_{01}$ mode profile. Left: cross-sectional side view, right: entrance view. The waveguide is a rectangular metallic waveguide with two identical quartz slabs loaded on the top and bottom.}
	\vspace*{-\baselineskip}
	\label{fig:DLW}
\end{figure}

Novel THz-driven structures are increasingly being considered as alternatives to conventional radio frequency accelerating structures due to potential for high accelerating gradients \cite{England,Hebling,Walsh}. Novel designs are also being explored for other uses such as removing energy chirp, bunch deflection, and beam diagnostics \cite{Fu,Pacey,Bettoni}. One promising design is the dielectric-lined waveguide (DLW). The use of cylindrical DLWs has been demonstrated for acceleration, with an energy gain of 7~keV for 60~keV electron bunches \cite{Nanni}. The structure considered in this work is a rectangular waveguide, with the longest sides each lined with a layer of dielectric of equal thickness, as shown in Figure \ref{fig:DLW}. The addition of dielectric, which couples the typical transverse electric and transverse magnetic modes to generate hybrid modes, Longitudinal Section Electric/Magnetic (LSM/LSE) \cite{Collin}, which for individual frequencies to propagate with a phase velocity, $v_p$, less than the speed of light, $c$. This results in positive interaction of a monochromatic wave with a propagating electron of velocity $v_e=v_p$, which is indefinite when there is no longitudinal acceleration. For deflection and acceleration, modes LSM$_{01}$ and LSM$_{11}$ respectively are excited. The polarization and spatial profile of the external field dictates the excited mode and therefore the effect of the DLW on a particle bunch, in this case electrons. For the purpose of this investigation the structure was considered for deflection only but it can also support accelerating modes. In this paper DLW design is discussed, and a general overview of THz-electron interaction in waveguide structures is provided with regards to interaction length. The analysis of interaction length and dispersion is valid for all structures in which the dispersion relation is modified such that $v_p = v_e$ for at least one frequency, provided $v_e$ remains constant. Coupling and its effects are ignored in this analysis. The effect of THz pulse length is demonstrated for the specific example of a DLW operating at 0.5~THz and designed to deflect 100~keV electrons, matching the energy of an existing electron gun. Future structures will be focussed on fully relativistic beams at the CLARA facility at STFC Daresbury Laboratory \cite{CLARA}. A comparison of DLWs and corrugated waveguides is presented with a focus on interaction length.
Transverse deflecting structures have uses in diagnostics; a time-dependent deflection of a bunch is projected onto a screen, converting longitudinal position to transverse position. 
\section{DLW design}
We consider a waveguide of length $L$ with dimensions given by $a$, $b$, and $w$, where $a$ is the half vacuum aperture, $b=a+\delta$, where $\delta$ is the dielectric thickness, and $w$ is the waveguide width. The presence of the dielectric modifies the waveguide dispersion relation from that of a hollow rectangular waveguide \cite{Xiao};
\begin{equation}
\label{eq:dispersionodd}
k_{y,mn}^b\tan\left(k_{y,mn}^b(b-a)\right) = \epsilon_r k_{y,mn}^a \cot\left(k_{y,mn}^a a\right)~,
\end{equation}
for $m=1,3,\ldots$, corresponding to a maximal longitudinal axial electric field $E_{z,0}$. For $m=0,2,\ldots$, corresponding to $E_{z,0}=0$, the dispersion relation is given by
\begin{equation}
\label{eq:dispersioneven}
k_{y,mn}^b\tan\left(k_{y,mn}^b(b-a)\right) = -\epsilon_r k_{y,mn}^a \tan\left(k_{y,mn}^a a\right)~,
\end{equation}
where $\epsilon_r$ is the relative permittivity of the dielectric. The wavenumbers directed along the stratification are given by
\begin{align}
k_{y,mn}^a&=\sqrt{k_0^2-\frac{m \pi}{w}^2-\beta_{mn}}~,\\
k_{y,mn}^b&=~\sqrt{\epsilon_r k_0^2-\frac{m \pi}{w}^2-\beta_{mn}}~,
\end{align}
in the vacuum aperture and dielectric respectively. $\beta_{mn}$ is the longitudinal wavenumber. Equations (\ref{eq:dispersionodd}) and (\ref{eq:dispersioneven}) must be solved numerically to find $\beta_{mn}$ and thus the dispersion relation. The choice of DLW geometry depends on the electron bunch dimensions, electron energy and type of interaction required. The vacuum aperture was limited to a minimum of 200~$\mu$m, double the transverse beam size of the planned interaction point. Quartz was selected as the dielectric due its availability and low loss tangent \cite{Hejase}. $w$, $a$, and $\delta$ were chosen by analysis of the effects of each parameter on the interacting bandwidth and axial voltage. It was found that matching bandwidth and maximising field was not possible; a compromise between the two was reached. Waveguide parameters were inserted into the dispersion relation which was solved analytically. Properties of the DLW such as field components were then analytically obtained from the calculated wavenumbers \cite{Xiao}, assuming the input power of each frequency is constant.  
\section{Interaction length}
The operating frequency $f_{op}$ is defined as that at which the phase velocity $v_p$ is equal to the longitudinal electron velocity, $v_e$. A monochromatic THz field at this frequency will propagate indefinitely with an electron without phase slippage. An electron will slip out of phase with a frequency other than $f_{op}$ over the length of the structure as $v_p \neq v_e$. Provided phase slippage is less than $\pm \pi/2$ respectively, the interaction will be entirely positive, giving the \textit{interacting bandwidth} constraint
\begin{equation}
\label{eq:propconst}
\left|\beta\left(\omega\right) - \beta_e(\omega)\right| L \leq \frac{\pi}{2}~,
\end{equation}
where $\beta\left(\omega\right) = \omega/v_p\left(\omega\right)$ is the longitudinal wavenumber, $\beta_e = \omega/v_e$ and $L$ is the structure length. Evaluating Equation (\ref{eq:propconst}) at $\omega_1$ and $\omega_2$, the frequencies at which phase slippage is $\pm \pi/2$ respectively and subtracting
\begin{equation}
\label{eq:propconst2}
\left(\beta\left(\omega_2\right)-\beta\left(\omega_1\right)-\frac{\left(\omega_2-\omega_1\right)}{v_e}\right)L_{int}=\pi~,
\end{equation}
giving an effective bandwidth of
\begin{equation}
\label{eq:bandwidthtrue}
\Delta f =\frac{\omega_2-\omega_1}{2\pi}=\frac{1}{2 \pi}v_e \left(\beta\left(\omega_2\right)-\beta\left(\omega_1\right) + \frac{\pi}{L_{int,p}}\right)~.
\end{equation}
This can be rearranged to get the interaction length $L_{int,p}$ for a given input THz pulse bandwidth, assuming that all frequencies are in phase at the start of the waveguide. Outside of this length, the edge frequencies interact negatively with the propagating electron. An alternative definition of the interaction length can be obtained by considering the slippage of the electron from the pulse envelope. For a transform-limited Gaussian pulse, which has the shortest pulse duration for a given optical spectrum, the time-bandwidth product is $\tau\Delta f \leq 0.44$. The interaction length, $L_{int,g}$, is calculated \cite{Healy} using
\begin{equation}
\label{eq:interactionlengthbandwidth}
\Delta f =\frac{0.44}{L_{int,g}\left(\frac{1}{v_g} - \frac{1}{v_e}\right)}~,
\end{equation}
assuming $v_g$ is frequency independent.  It must be emphasised that both equations are approximations; Equation (\ref{eq:bandwidthtrue}) only considers individual frequencies, not a real pulse, and Equation (\ref{eq:interactionlengthbandwidth}) assumes no group velocity dispersion.
\section{THz-electron interaction}
\begin{figure*}[h]
	\centering	
	\begin{subfigure}{0.4\textwidth}
		\centering
		\includegraphics[width=1\linewidth]{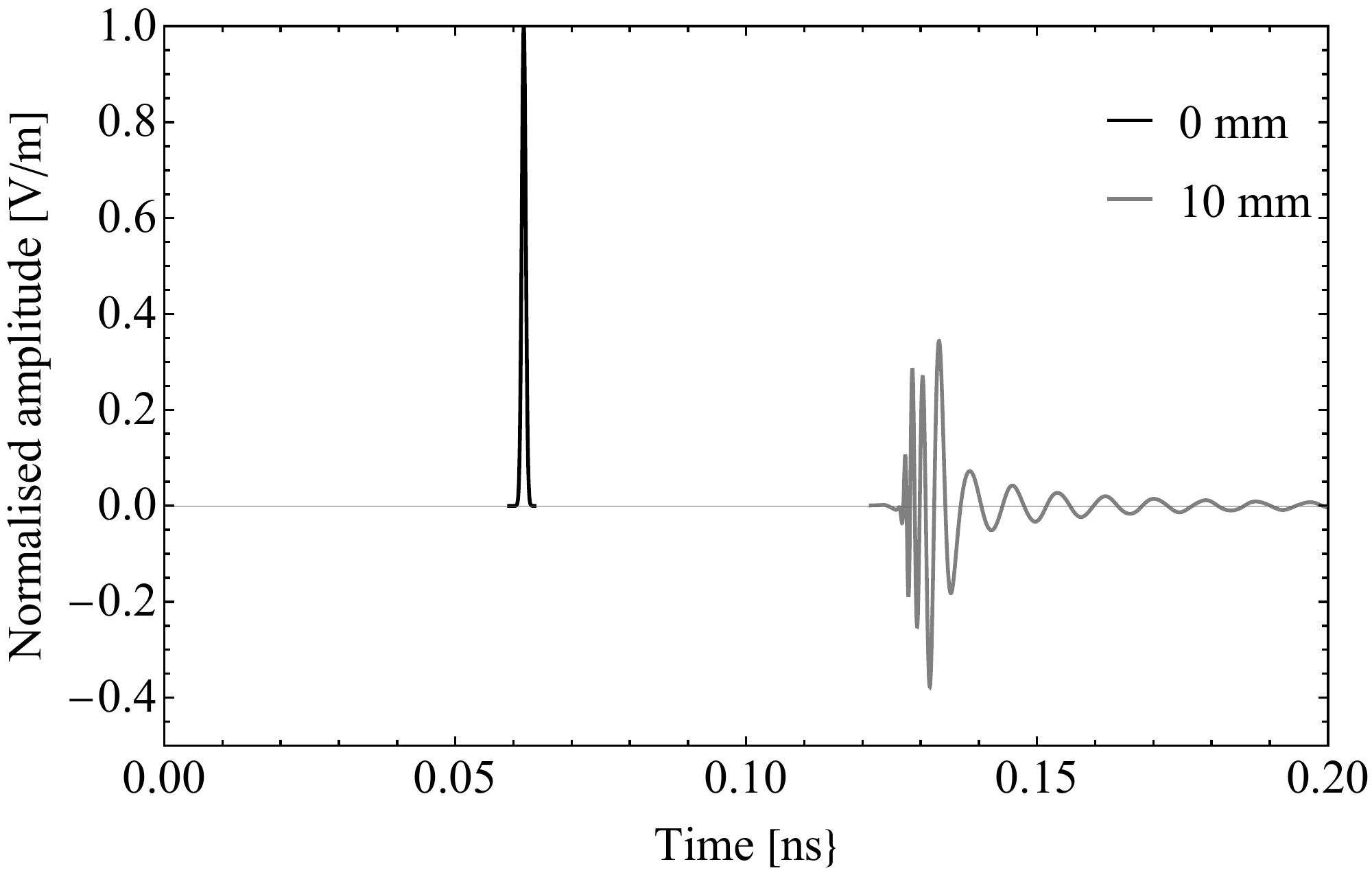}
		\caption{}
		\label{fig:singlepulseamp}
	\end{subfigure}%
\hspace*{20mm}%
	\begin{subfigure}{0.4\textwidth}
		\centering
		\includegraphics[width=1\linewidth]{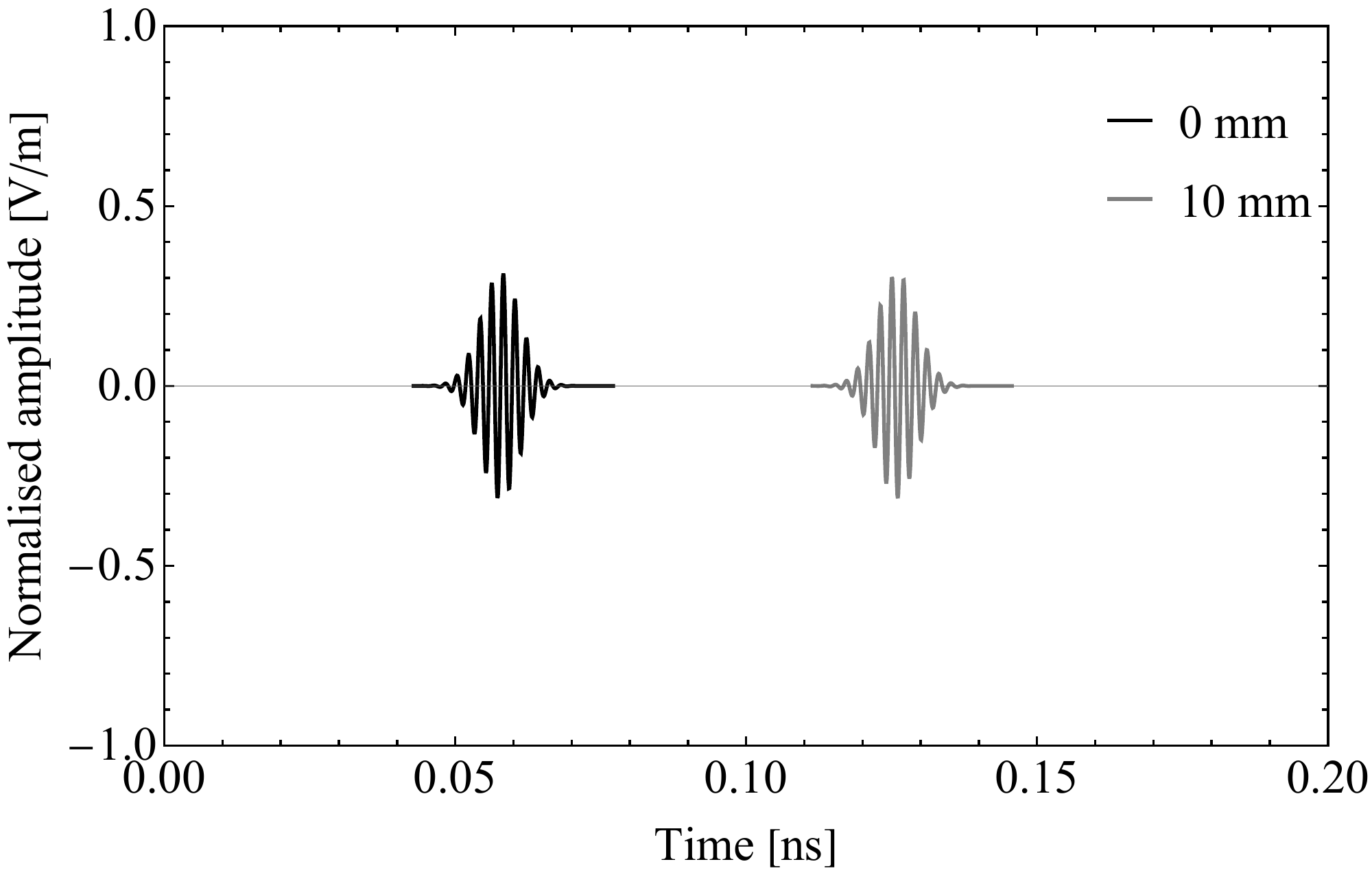}
		\caption{}
		\label{fig:tenpulseamp}
	\end{subfigure}
	\caption{Dispersion of a THz pulse due to DLW. The amplitude of $E_y$ field on-axis is measured as a function of time. a) single-cycle THz pulse, b) ten-cycle THz pulse. The input power of the THz pulse across the input face of the waveguide is the same in both cases. And the amplitude is normalised to the input single-cycle pulse.}
	\label{fig:amp}
\end{figure*}

\begin{figure*}[h]
	\centering
	\begin{subfigure}{0.4\textwidth}
		\centering
		\includegraphics[width=1\linewidth]{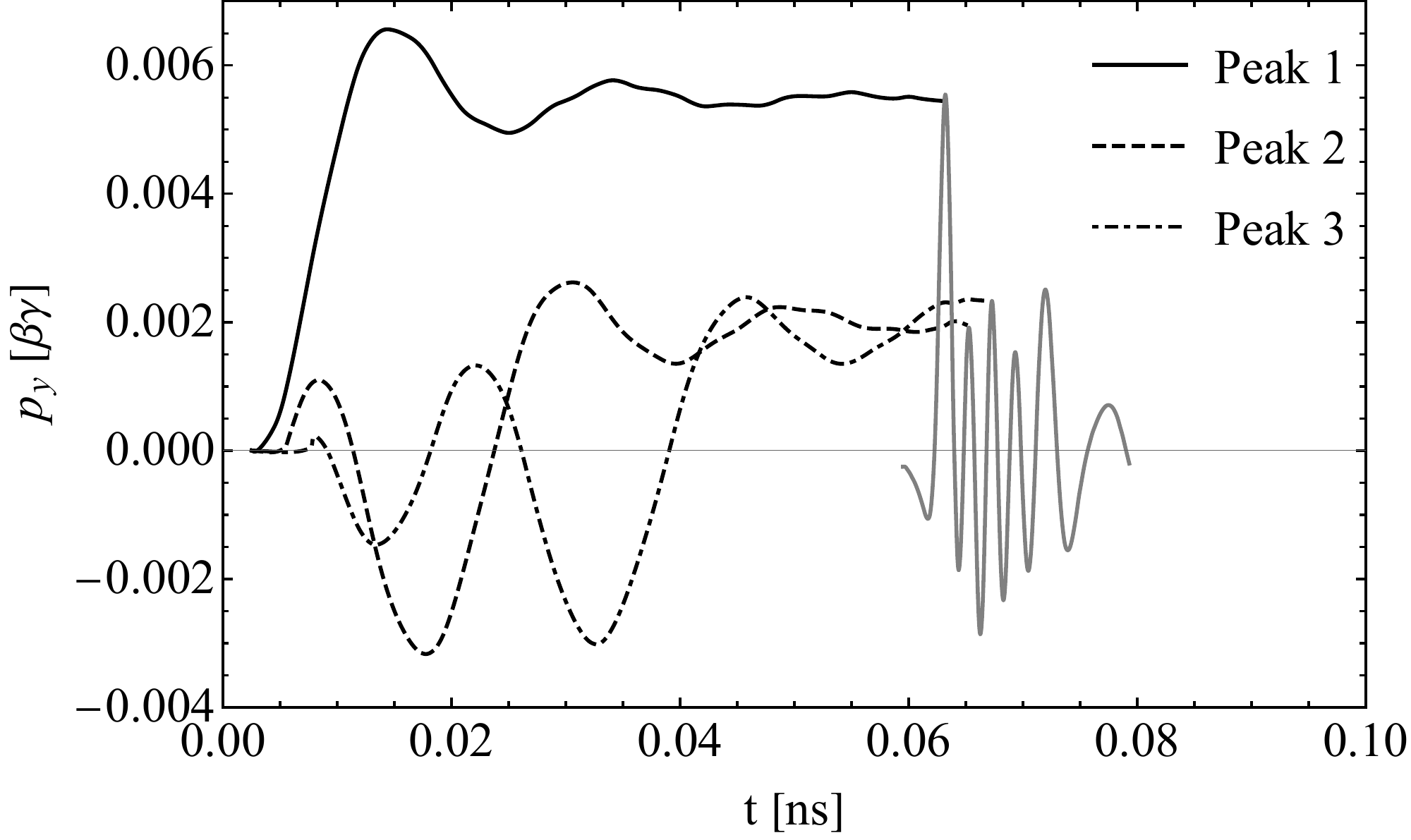}
		\caption{}
		\label{fig:singlecycleymom}
	\end{subfigure}%
\hspace*{20mm}%
	\begin{subfigure}{0.4\textwidth}
		\centering
		\includegraphics[width=1\linewidth]{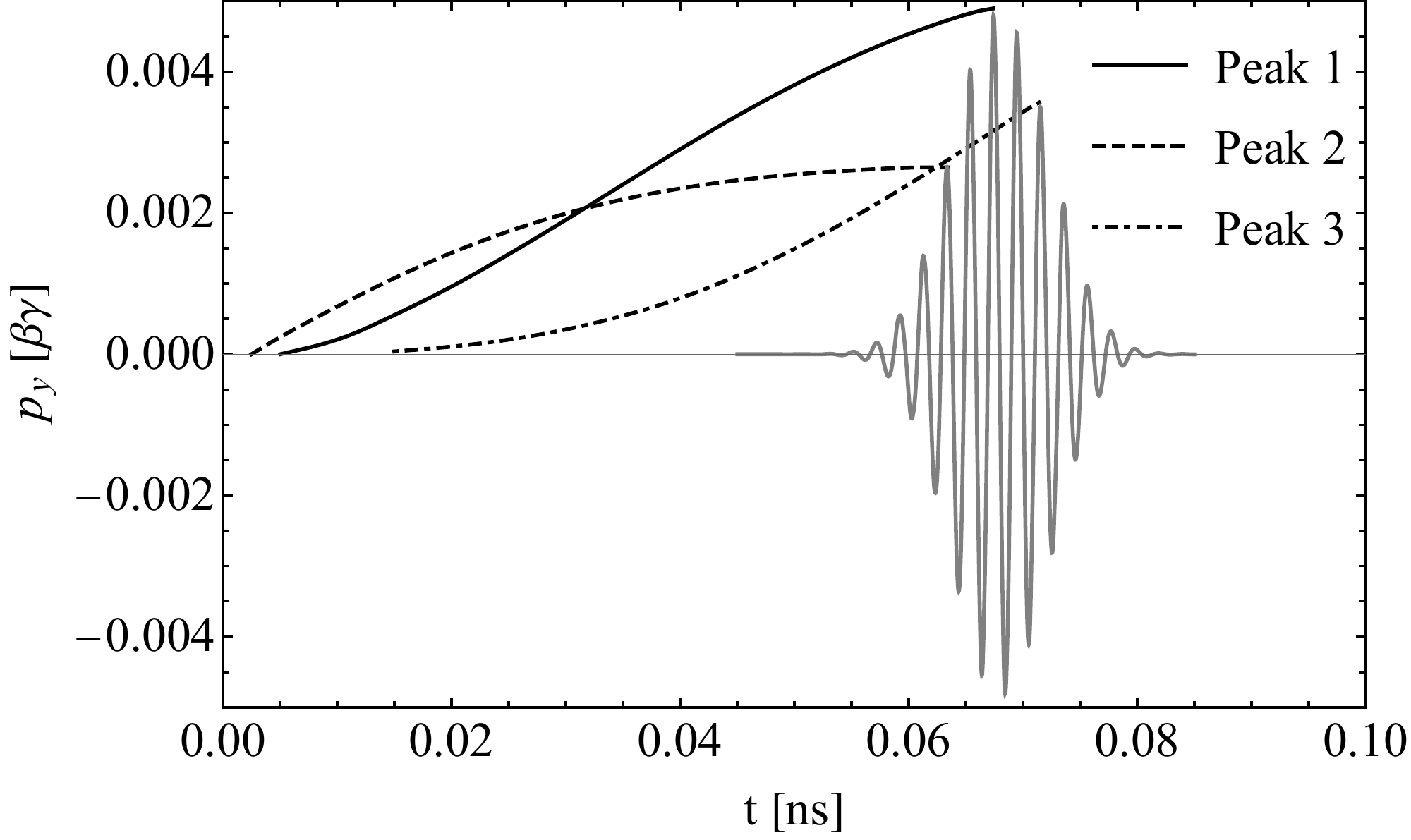}
		\caption{}
		\label{fig:tencycleymom}
	\end{subfigure}
	\caption{Grey: Momentum kick experienced by on-axis electrons due to interaction with a THz pulse as a function of injection time at the end of the 10~mm structure. Black, blue, red: following a single electron on a momentum peak from start to end of the waveguide. a) single-cycle THz pulse, b) ten-cycle THz pulse.}
	\label{fig:mom}
\end{figure*}

The THz-electron interaction was simulated using the CST Particle-in-Cell (PIC) solver \cite{CST}. Two THz pulses were investigated; a single-cycle Gaussian with bandwidth 0-1~THz and a ten-cycle pulse with bandwidth 0.3-0.7~THz, corresponding to planned experimental parameters. A 73~ps bunch of on-axis probe particles with no initial transverse emittance was launched to co-propagate with the THz pulse; only single particle dynamics were considered. Each pulse was launched directly into the LSM$_{01}$ mode. Figure \ref{fig:amp} shows the normalised axial $E_y$ field of the two pulses just after launch and after propagating 10~mm into the waveguide. The input power of the THz pulse across the input face of the waveguide is the same in both cases. The single-cycle pulse was highly dispersed by waveguide propagation and the maximum amplitude decreased as a result, whereas the shape of the 10-cycle pulse remained largely unchanged. The effect of the two THz pulses on interacting electrons with different injection times is shown in Figure \ref{fig:mom}. 
The interaction with a ten-cycle pulse is shown in Figure \ref{fig:tencycleymom}. An electron injected with the front of the pulse, such as that tracked to `Peak 2', slips out of the pulse as $v_g \neq v_e$ and so has a lower y-momentum gain than `Peak 1', corresponding to the electron with maximal gain. This was injected towards the end of the THz pulse on a peak. Electrons on `Peak 1' and `Peak 3' are still interacting as they have not propagated completely through the pulse by the end of the structure. Therefore a 10-cycle pulse would provide a greater maximum interaction given a longer structure.In the case of a single-cycle pulse, electrons injected just behind the trailing edge of the single-cycle pulse are given the largest kick in $y$, corresponding to `Peak 1' in Figure \ref{fig:singlecycleymom}. Later electrons, such as those with a final energy corresponding to `Peak 2' and `Peak 3' experience less of a kick.  The three particles selected for investigation leave the 10~mm structure with maximised net deflection but experience a periodic variation in interaction as a result of pulse dispersion. The periodicity is approximately 0.02~ns, with a positive interaction over 0.01~ns, corresponding to a length of 1.6~mm. From Equation (\ref{eq:bandwidthtrue}), the interaction length should be $\approx700~\mu$m. The simulated interaction is more than doubled for an electron injected with the peak of the THz pulse. This suggests that our definition of interaction length is an underestimate as the interaction of the central frequencies in the bandwidth is not cancelled out by the edge frequencies. It is therefore preferable to minimise dispersion, which is achieved via the use of a narrowband THz pulse, although practically this is at the expense of THz power as narrowband sources are typically less efficient than broadband sources.
\section{Comparison to corrugated waveguides}
\begin{figure}[h]
	\centering
	\includegraphics[width=0.8\linewidth]{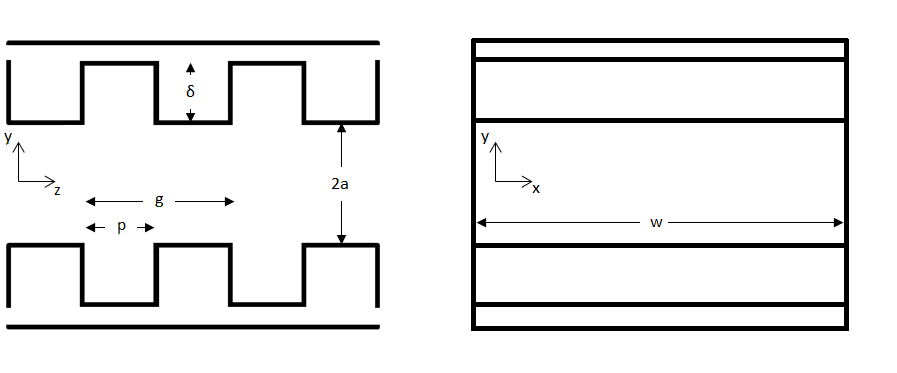}
	\caption{Cross-sectional view of the corrugated waveguide. Left: side view, right: entrance view. The waveguide is a rectangular metallic waveguide with periodic corrugations on the top and bottom.}
	\vspace*{-\baselineskip}
	\label{fig:corr}
\end{figure}

\begin{figure}[h]
	\centering
	\includegraphics[width=0.8\linewidth]{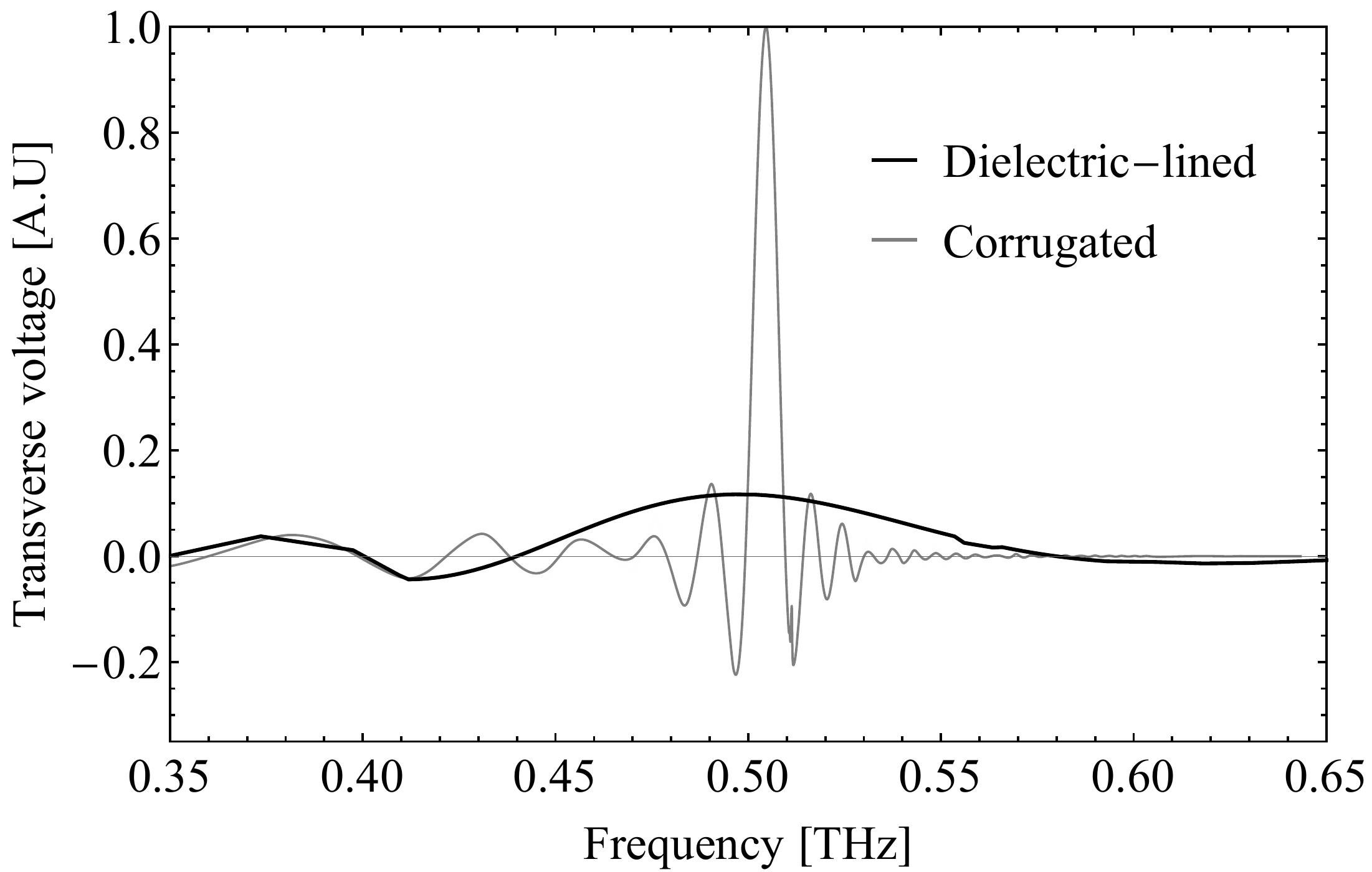}
	\caption{Transverse voltage $V_y$ of the corrugated waveguide and DLW as a function of frequency, normalised to DLW maximum. The power is the same for each input frequency.}
	\vspace*{-\baselineskip}
	\label{fig:Vt}
\end{figure}

\begin{figure}[]
	\centering
	\includegraphics[width=0.8\linewidth]{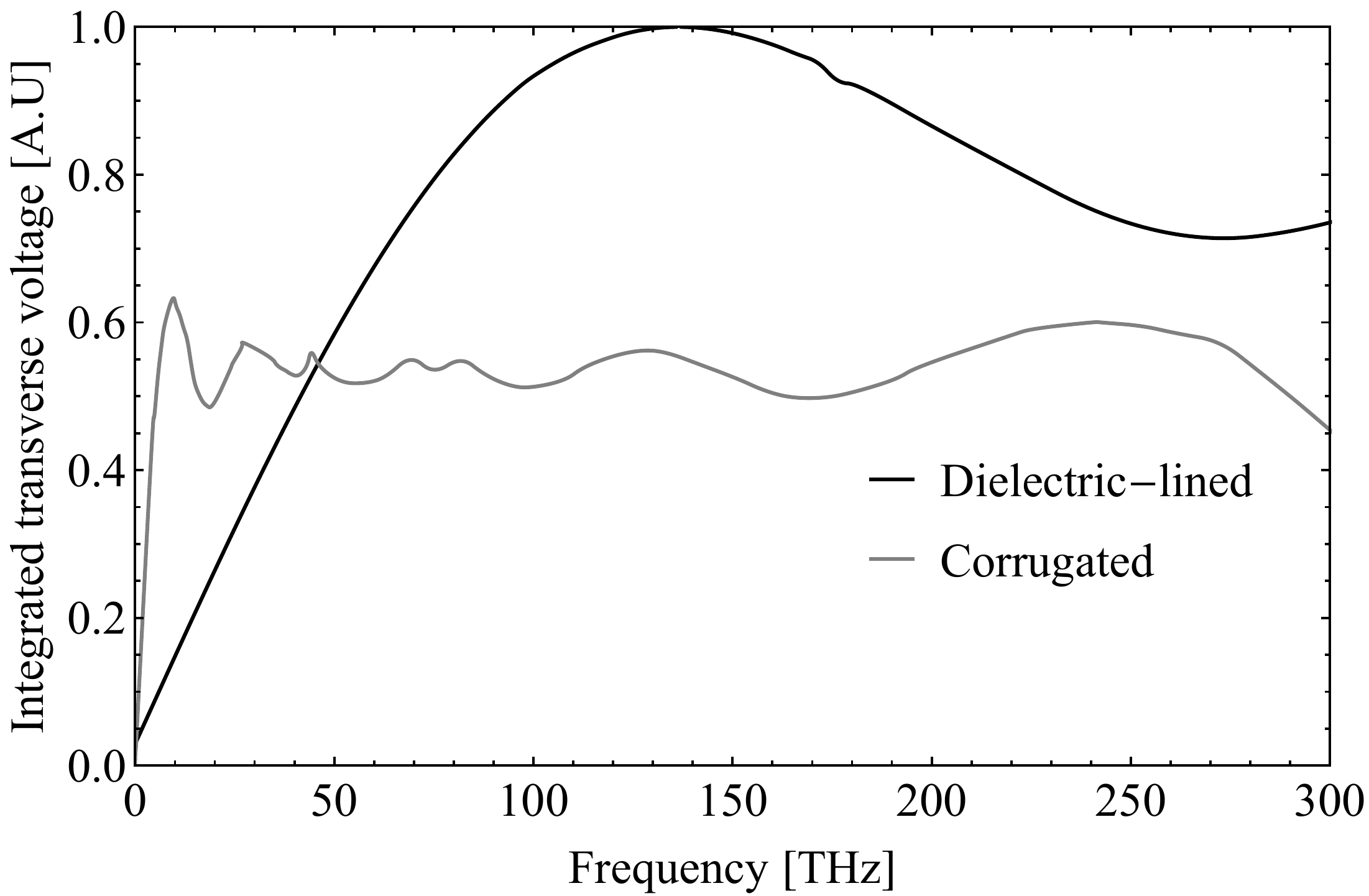}
	\caption{Transverse voltage $V_y$ of the corrugated waveguide and DLW, integrated over a given bandwidth and normalised to the DLW maximum value. The power spectrum over each bandwidth is constant.}
	\vspace*{-\baselineskip}
	\label{fig:Vtbandwidth}
\end{figure}

A corrugated rectangular metallic waveguide, such as shown in Figure \ref{fig:corr}, is a simple alternative to the DLW. It is similar to a conventional iris-loaded cavity in that longitudinal variation in the aperture serves to modify the dispersion relation from that of a metallic rectangular waveguide. The dispersion relation and electromagnetic fields were calculated using CST Microwave Studio (MWS).
A comparison between the two structure types was performed for a structure optimized to deflect 100~keV electrons, primarily by assessing the phase and group velocity and the transverse voltage,
\begin{equation}
V_y\left(\omega\right)=\int^L_0 \left(E_y\left(\omega\right)+v_e B_x\left(\omega\right)\right)\exp\left[-i \left(\beta\left(\omega\right)z-\frac{\omega}{v_e} z\right)\right]\mathrm{d}z~,
\label{eq:Vt}
\end{equation}
where only the $y$ direction is considered as there is no axial field acting in $x$. The electron velocity is included to account for the variation of $v_p$ with frequency, as $v_p=\omega/\beta \left(\omega\right)$. The optimised structure dimensions are shown in Table \ref{tab:nonreldeflcorr} for an operating frequency $f_{op}=0.5$~THz and electron energy 100~keV. 
\begin{table}[h]
	\centering
	\caption{Parameters of the optimised corrugated and dielectric-lined waveguides. Parameters are defined in Figure \ref{fig:corr}.}
	\label{tab:nonreldeflcorr}
	\begin{tabular}{c|c|c}
		Parameter & Corrugated & DLW \\\hline
		a~($\mu$m)   & 100 & 100   \\
		w~($\mu$m)  & 1000 & 1000   \\
		p~($\mu$m)   & 35 & -   \\
		g~($\mu$m)   & 40 &-   \\
		$\delta$~($\mu$m)   & 108.5 &242\\
		$\epsilon_r$ & - & 3.75 \\  
	\end{tabular}
\end{table}
\begin{figure}[h]
	\centering
	\includegraphics[width=0.8\linewidth]{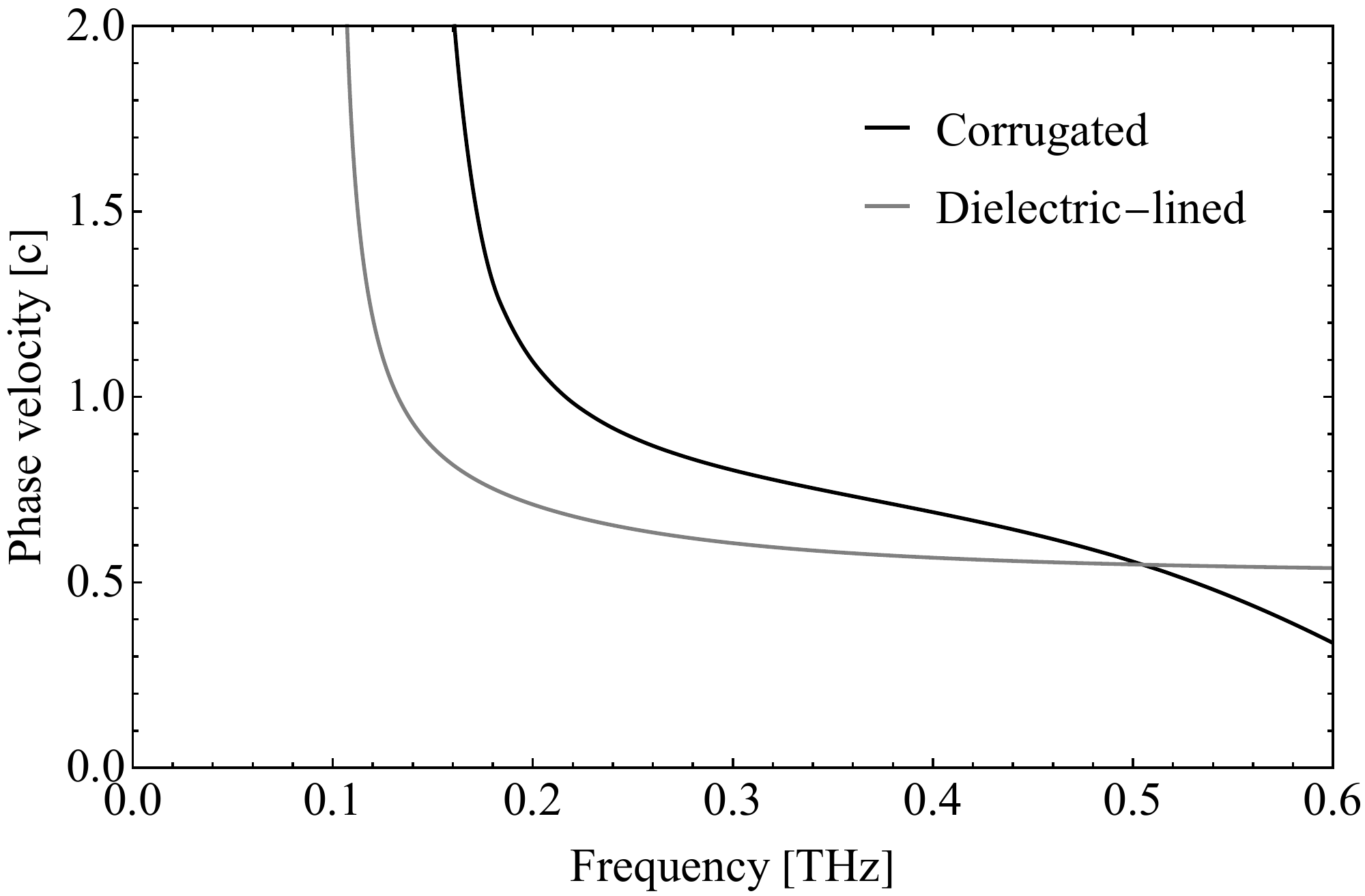}
	\caption{Phase velocity of the corrugated waveguide and DLW as a function of frequency. The crossing point at 0.5~THz is when $v_p=v_e$.}
	\vspace*{-\baselineskip}
	\label{fig:vp}
\end{figure}
\begin{figure}[h]
	\centering
	\includegraphics[width=0.8\linewidth]{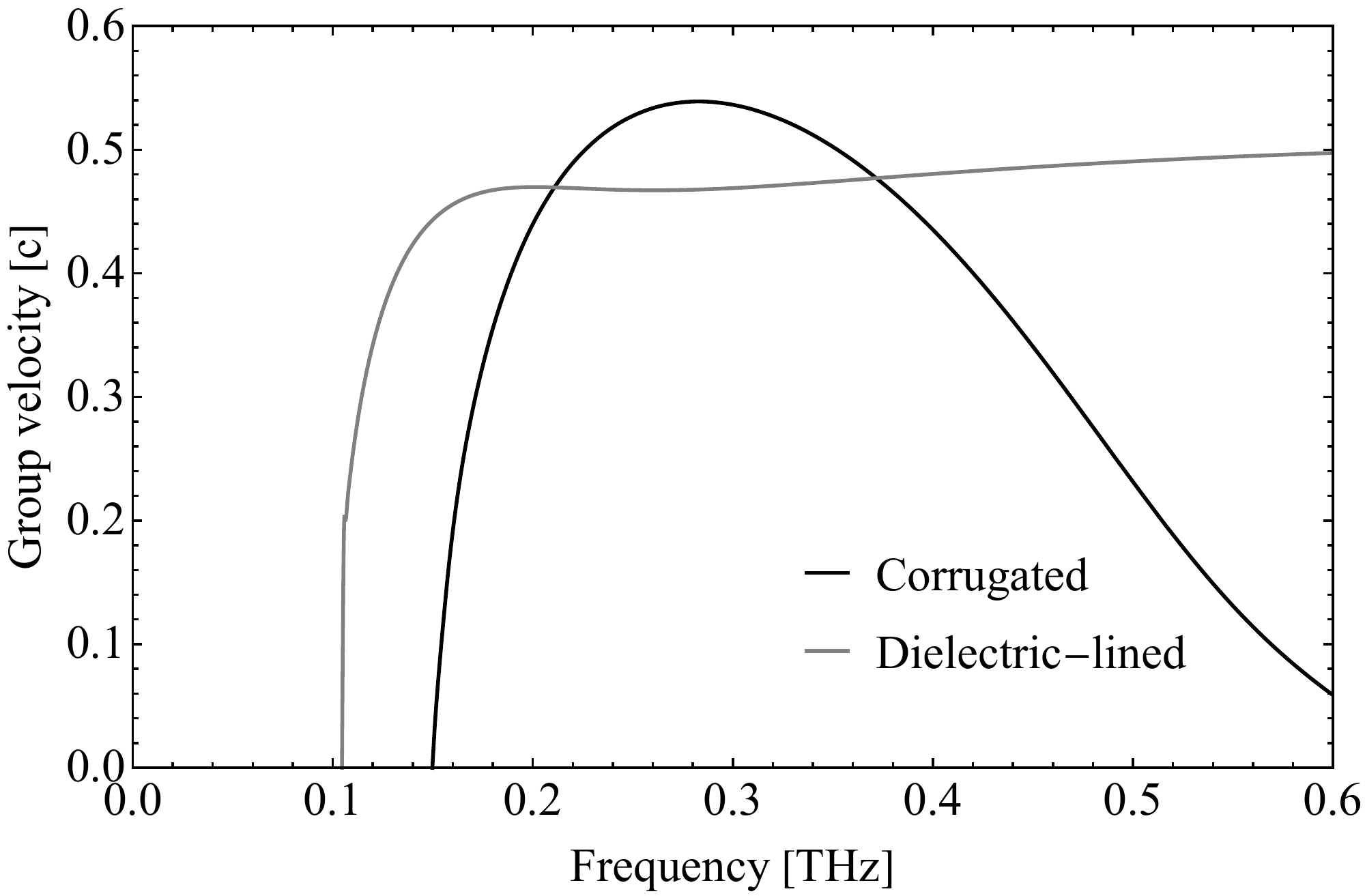}
	\caption{Group velocity of the corrugated waveguide and DLW as a function of frequency. Neither waveguide achieves $v_g=v_e$ at 0.5 THz, necessary for indefinite interaction.}
	\vspace*{-\baselineskip}
	\label{fig:vg}
\end{figure}
Figures \ref{fig:vp} and \ref{fig:vg} show that the DLW is less dispersive than the corrugated waveguide as both $v_p$ and $v_g$ vary less close to $f_{op}$. In both cases $v_g < v_e$ for all frequencies and so the pulse envelope propagates slower than the electron bunch, limiting the interaction length. Using Equation (\ref{eq:bandwidthtrue}), a corrugated structure of length 10~mm requires a input THz bandwidth less than 5.2~GHz for monotonic interaction, whereas Equation (\ref{eq:interactionlengthbandwidth}) estimates 4.7~GHz, with the discrepancy arising due to assumptions in both equations. For the DLW, Equations (\ref{eq:bandwidthtrue}) and (\ref{eq:interactionlengthbandwidth}) give $\Delta f$ of 70~GHz and 62~GHz respectively. Figure \ref{fig:Vt} shows the transverse voltage $V_y$ in both cases, calculated using data taken from CST for individual frequencies, assuming power of each frequency is constant, and inserted into Equation (\ref{eq:Vt}). The corrugated waveguide has a higher $V_y$ at the centre frequency than the DLW, but $V_y$ for other frequencies is lower and periodicity of variation is shorter. Figure \ref{fig:Vtbandwidth} shows $V_y$ integrated over a given frequency range $\Delta\omega$, $V_y^{int}=\int_{\omega_0+\Delta \omega/2}^{\omega_0-\Delta \omega/2}V_y\left(\omega\right)$, assuming a constant spectral distribution and centred on $\omega_0$. With increasing bandwidth, the power of the pulse increases. This shows that the use of a DLW is favourable except for bandwidths less than $\sim$50~GHz. Despite the larger maximum $V_y$ of the corrugated waveguide, integrated voltage is smaller as a result of the larger $v_p$,~$v_g$ variation. The bandwidth corresponding to maximum $V_y$ was 9.3~GHz for the corrugated waveguide and 132~GHz for the DLW.  The approximations made in Equations (\ref{eq:bandwidthtrue}, \ref{eq:interactionlengthbandwidth}) are therefore not valid in this analysis.
\section{Conclusions}
A DLW structure, designed to deflect 100~keV electrons, has been investigated and compared with a corrugated waveguide. The effect of THz pulse bandwidth on magnitude of transverse deflection has been numerically investigated and it has been established that choice of THz pulse bandwidth is important to maximise deflection. If bandwidth is too small, an electron will not pass through the entire pulse and so will not experience maximum possible deflection. A THz pulse with a large bandwidth is dispersed and the pulse peaks do not propagate with the same velocity as the electron, resulting in variation of transverse kick direction. The ideal pulse bandwidth results in an electron passing through the entire pulse over the length of the structure, starting at the trailing edge of the pulse and exiting the waveguide just ahead of the leading edge. For the corrugated waveguide considered here, the large dispersion leads to a requirement of a 9.3~GHz pulse bandwidth compared to the 132~GHz bandwidth of the DLW over a 10~mm structure. The higher peak field strengths achievable with short broadband THz pulses, together with increased efficiency of non-linear generation, leads to more favourable conditions for DLW structures. Future experimental work will investigate this. Inclusion of group velocity dispersion and higher-order effects will be included in future work by time-domain analysis of THz-electron interaction, thus investigating the effect of real THz pulses. 

%
%
%

\end{document}